 \documentclass{aa}
\usepackage{graphicx}
\usepackage{longtable,lscape}
\usepackage{natbib}
\bibpunct{(}{)}{;}{a}{}{,}

\def\H0 {$H_{\rm o}$}

\def\solmass {\hbox{M$_{\odot}$}}

\def\lsols {\hbox{L$_{\odot}$}}

\def\numd {\hbox{$n\,({\rm H}_2$) }}

\def\percc {$\hbox{{\rm cm}}^{-3}$}    

\def\NH3 {\hbox{${\rm NH}_{3}$}}                  
\def\CH3C2H {\hbox{${\rm CH}_3{\rm C}_2{\rm H}$}} 
\def\CH3OD {\hbox{${\rm CH}_3{\rm OD}$}}          
\def\HC15N {\hbox{${\rm HC}^{15}{\rm N}$}}        
\def\HN13C {\hbox{${\rm HN^{13}C}$}}              
\def\HCOP {\hbox{${\rm HCO}^+$}}                  
\def\SO2 {\hbox{${\rm SO_{2}}$}}                  
\def\H2S {\hbox{${\rm H_{2}S}$}}                  
\def \ga{\mathrel{\mathchoice   {\vcenter{\offinterlineskip\halign{\hfil
$\displaystyle##$\hfil\cr>\cr\sim\cr}}}
{\vcenter{\offinterlineskip\halign{\hfil$\textstyle##$\hfil\cr
>\cr\sim\cr}}}
{\vcenter{\offinterlineskip\halign{\hfil$\scriptstyle##$\hfil\cr
>\cr\sim\cr}}}
{\vcenter{\offinterlineskip\halign{\hfil$\scriptscriptstyle##$\hfil\cr
>\cr\sim\cr}}}}}

\begin{document}

%
%
\title{Evolution of the ISM in Luminous IR galaxies}
%
\author{W.A.~Baan\inst{1}, A.F.~Loenen\inst{2},  \and M.~Spaans\inst{3}}
\offprints{W.A. Baan, \email{baan@astron.nl}}

\institute{
 ASTRON, Oude Hoogeveensedijk 4, 7991 PD Dwingeloo, The Netherlands 
 \and
Sterrewacht Leiden, P.O. Box 9513, 2300 RA Leiden, The Netherlands
\and
 Kapteyn Astronomical Institute, P.O. Box 800, 9700 AV Groningen, The Netherlands
}
\date{Received date 1/04/2009; accepted date 7/04/2010}
\authorrunning{Baan et al.}
\titlerunning{Evolution of the ISM in luminous IR galaxies}
\abstract {} {Molecules that trace the high-density regions of the
  interstellar medium may be used to evaluate the changing physical
  and chemical environment during the ongoing nuclear activity in
  (Ultra-)Luminous Infrared Galaxies.}  {The changing ratios of the
  HCN\,(1$-$0), HNC\,(1$-$0), \HCOP\,(1$-$0), CN\,(1$-$0) and
  CN\,(2$-$1), and CS\,(3$-$2) transitions were compared with the
  HCN\,(1$-$0)/CO\,(1$-$0) ratio, which is proposed to represent the
  progression time scale of the starburst. These diagnostic diagrams
  were interpreted using the results of theoretical modeling using a
  large physical and chemical network to describe the state of the
  nuclear ISM in the evolving galaxies.}  {Systematic changes are seen
  in the line ratios as the sources evolve from early stage for the
  nuclear starburst (ULIRGs) to later stages. These changes result
  from changing environmental conditions and particularly from the
  lowering of the average density of the medium. A temperature rise
  due to mechanical heating of the medium by feedback explains the
  lowering of the ratios at later evolutionary stages.  Infrared
  pumping may affect the CN and HNC line ratios during early
  evolutionary stages.}  {Molecular transitions display a behavior
  that relates to changes of the environment during an evolving
  nuclear starburst. Molecular properties may be used to designate the
  evolutionary stage of the nuclear starburst. The HCN(1$-$0)/CO(1$-$0) 
  and HCO$^{+}$(1$-$0)/HCN(1$-$0) ratios serve as indicators
  of the time evolution of the outburst.

  }

\keywords{infrared: galaxies -- ISM: molecules -- radio line:
galaxies -- galaxies: active -- galaxies: starburst, -- Masers}
\maketitle
\section{Introduction}
 
The (Ultra-) Luminous Infrared Galaxies (ULIRGs) are powered by a
(circum-)nuclear starburst and/or an Active Galactic Nucleus
(AGN). ULIRGs belong to a much larger population of Sub-Millimeter
Galaxies (SMGs) that peaks at redshift 2$-$3 and extends to z =
$\sim$6.  The relatively short episodes of intense nuclear activity in
ULIRGs and SMGs have likely been triggered by galaxy mergers or
collisions, which produce the most luminous galaxies in the universe.

Among other spectral observations, the emissions from the high-density
molecular material in the nuclear region of external galaxies allow a
clear view of the ISM and are a preferred tool for diagnosing the
physical and chemical environment of the starforming and AGN-excited
ISM. A strong relation has been found between the far-infrared
luminosity and the luminosity of molecular emissions, that are
indicators of the high-density (\numd$\ga$10$^{4-6}$\,\percc)
component of the interstellar medium \citep{GaoS2004a,CurranAB2000,
 BaanEA2008}. Since these integrated emissions predominantly arise
from the nuclear region, they are closely associated with the nuclear
starburst/AGN environment and the production of the FIR and
sub-millimeter luminosities.

Each molecular species responds differently to the changing physical
and chemical environment of the nuclear region. The collective
behavior of molecular tracers may thus be used to characterize the
evolution of the nuclear environment during a nuclear activity.  In
this paper, we consider evolutionary changes of the ISM as seen by the
molecular tracer emissions and resulting from the ongoing
star-formation, the AGN activity, and any feedback
processes. Diagnostic interpretation of the behavior of molecular line
ratios is based on theoretical modeling using a large physical and
chemical network to describe the state of the nuclear ISM in the
evolving galaxies.

This study consider mostly single molecular transitions for the
multi-molecular modeling.  The inclusion of multiple higher
transitions will augment the diagnostic accuracy of the modeling
analysis. Modeling of the observed ground state transitions already
predicts the behavior of higher transitions and their ratios and will
be used in future work to incorporate newly obtained observational
data and data expected from Herschel and ALMA.

\section{Modeling the Evolving ISM}

The emissions of molecular species observed in active nuclei are
determined by their chemical abundances, the density and temperature
of the medium, their column density, and the excitation conditions in
the ISM. While multiple level studies of single molecules may be used
to reveal the density and/or kinetic temperature of ISM components,
the ensemble of emissions of tracer molecules will diagnose the
dominant physical and chemical properties of the environment.

\begin{figure}[t!]
\includegraphics[width=9.2cm,clip]{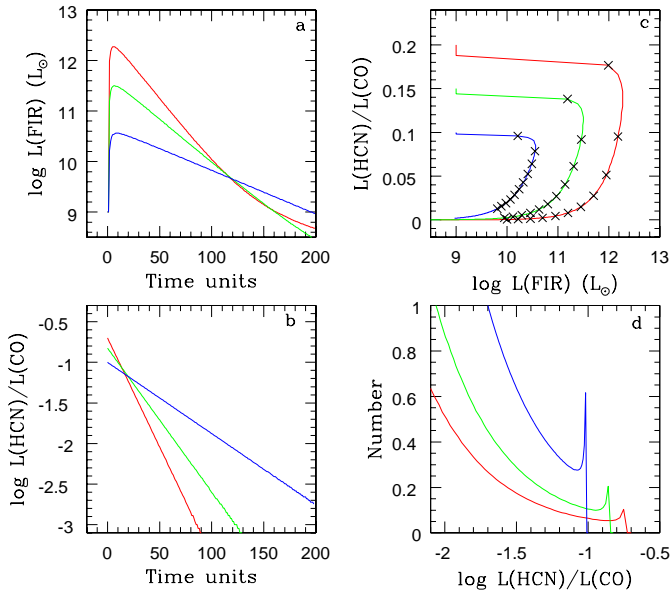}
\caption{Simulations of a starburst. Three initial conditions are
  considered (see text) and the curves are color coded. {\bf a)} The
  variation of the FIR-luminosity with representative time units of 5
  x 10$^{5}$ yr.  {\bf b)} The constant percentage-wise depletion of
  the high-density (HD) component relative to the low-density (LD)
  component. {\bf c)} The depletion of the high-density component as a
  function of the FIR-luminosity.  The crosses indicate intervals of
  10 time-steps. {\bf d)} The predicted number distribution of
  star-bursting FIR galaxies as a function of the HD/LD emission line
  ratio. The narrow spikes occur at the peak of the
  FIR-luminosity.} \label{simul}
\end{figure}

\subsection{Modeling the Evolution of a Starburst}
\label{sec:evolution}

The observed nearly linear relations between the emission luminosity
of high-density tracers and the far-infrared luminosity (FIR)
represent the response of the nuclear medium to the activity in the
evolving nucleus \citep{GaoS2004a,CurranAB2000, BaanEA2008}.  The
star-formation that generates the FIR luminosity also proportionally
depletes and destroys the high-density (HD) component of the nuclear
ISM, and the HD line emission decreases in time.  The low-density (LD)
component, however, is not involved in the star-formation process and
will remain (largely) the same. For this reason, the changing HD/LD
ratio during a nuclear outburst may serve as an indicator of the
evolutionary age of the outburst.  In section \ref{sec:indicators}, it
will be proposed to use the HCN\,(1$-$0)/CO\,(1$-$0) line emission
ratio as an indicator of the HD/LD emission ratio and the evolutionary
stage of the starburst.

The time evolution of the FIR luminosity of a nuclear starburst
may be characterized by a rapid rise to a peak luminosity followed
by an exponential tail \citep{LoenenBS2006,BaanEA2008}.  In order to
obtain further understanding about the FIR evolution and the
HD-depletion during a starburst, we consider a simple simulation
where during each time interval a fixed depletion fraction
$f_{dep}$ of the high-density medium is used and destroyed by
the star-formation process. The HD mass component used for
star-formation diminishes steadily with each timestep as:

\begin{equation} dM_{HD}(t) = M_{HD}(t-1) * f_{dep}. \label{eq:1} \end{equation}

\noindent The FIR-luminosity follows from a time-delayed
(diffusion-like) response from each of the mass fractions used in the
star-formation process during earlier time intervals.  This function
may be defined as:

\begin{equation} L_{\rm FIR}(t) = L_{\rm FIRm}\, \sum^{t}_{t_{i}=0}\, dM_{HD}(t_i)\,
\left(\frac{T}{t-t_i}\right)^{2.5} e^{-T/(t-t_i)}, \label{eq:2}
\end{equation}

\noindent where $L_{\rm FIRm}$ is the maximum luminosity and
$T$ is the characteristic luminosity diffusion timescale of the burst.

The FIR-luminosity trails the mass-consumption in time, which is
consistent with reality where lower mass stars have a delayed impact
on the FIR-luminosity. This simulation differs from the one presented
earlier in \cite{BaanEA2008} because here the depletion of the HD
material is used to determine the FIR-luminosity, while previously the
HD-depletion was deduced from the magnitude of the FIR-luminosity.

The decaying $L_{FIR}$ luminosity curve is determined by the peak
luminosity, the HD depletion rate per time interval, and the
delay-time-scale $T$ for the production of the FIR luminosity.  The
starburst simulations depicted in Fig.~\ref{simul} are for three
star-formation scenarios, respectively red, green, and blue curves,
with initial values for the HD/LD ratio (0.2, 0.15, and 0.1), the
depletion fraction $f_{dep}$ per timestep (6\%, 4\% and 2\%), and the
peak luminosity (10$^{12.3}$, 10$^{11.5}$, and 10$^{10.6}$ \lsols).
As mentioned before, the mass of the low-density component is assumed
to remain unchanged. The luminosity diffusion timescale of the burst
$T$ is set equal to 2.5 timesteps.  Assuming a 10$^{8}$ yr duration
for the outburst, each time-step in the simulation has an approximate
duration of 5 x 10$^{5}$ yr. While the scaling of the number and the
FIR-luminosity is still arbitrary, the curves may be compared with
observations.

The predicted behavior of the simulated starbursts is presented in the
graphs of Fig.~\ref{simul}. They (expectedly) show that more rapid
HD-consumption (red curve in frame b) also produces a more pronounced
luminosity curve and a faster luminosity decay (frame a).  The
continuing reduction of the HD/LD is plotted as function of
FIR-luminosity in frame c).  Because of the decreasing HD consumption
over time and the delay in the FIR production, the predicted number of 
sources in Frame d) increases towards lower values of $L_{\rm HD}$/$L_{\rm LD}$.
The diagrams in frames c) and d) may be compared with observational data 
in order to test the validity of the modeling, as presented in Section 
\ref{sec:indicators} below.

\begin{table}[t!]
\caption{Critial Density for Molecular Transitions}
\label{crit}
\begin{tabular}{lll}
\hline \hline
\noalign{\smallskip}
Molecule & Transition Designation & Critical Density  \\
                  &               &  (cm$^{-3}$)         \\
\hline 
\noalign{\smallskip}
CO    & J = 1$-$0                 & $2.0\times 10^3$  \\     
CO    & J = 2$-$1                 & $1.4\times 10^4$  \\     
\HCOP & J = 1$-$0                 & $2.2\times 10^5$  \\     
CN    & N = 1$-$0, J = 3/2$-$1/2  & $4.0\times 10^5$  \\         
CS    & J = 3$-$2                 & $1.4\times 10^6$  \\
HCN   & J = 1$-$0                 & $3.1\times 10^6$  \\
HNC   & J = 1$-$0                 & $3.9\times 10^6$  \\
CN    & N = 2$-$1, J = 5/2$-$3/2  & $1.9\times 10^7$  \\
\hline 
\noalign{\smallskip}
\end{tabular}
\newline Note: Based on data from LAMDA database for a kinetic 
temperature of 60-70 K \citep{SchoierEA2005}.
\label{table1}
\end{table}

\subsection{Chemical Evolution of the Starburst}
\label{sec:chem-evol-starb}

The changing properties of the evolving nuclear environment trigger
many related and simultaneous processes. After some high-density
material is used to form the first generation of stars, the remaining
high-density component will increasingly be affected by these stars,
their UV-radiation fields, and the resulting (ultra-) compact HII
regions. After the first massive stars produce supernovae and
remnants, the dissipation of mechanical energy of the shocks will
further raise the temperature and disperse the surrounding
high-density material.  Such processes will modify the formation of
subsequent generations of stars and give a continually changing ISM.
It should be noted that the time scales for shocks and
photo-dissociation by UV-radiation are very short \cite[10$^{4}$ $-$
10$^{5}$ yr;][] {BerginEA1998} as compared with the time scale of a
standard starburst \cite[10$^{7}$ $-$ 10$^{8}$ yr;][]{Coziol1996}.

\begin{figure}[t!]
\includegraphics[width=9.0cm,clip]{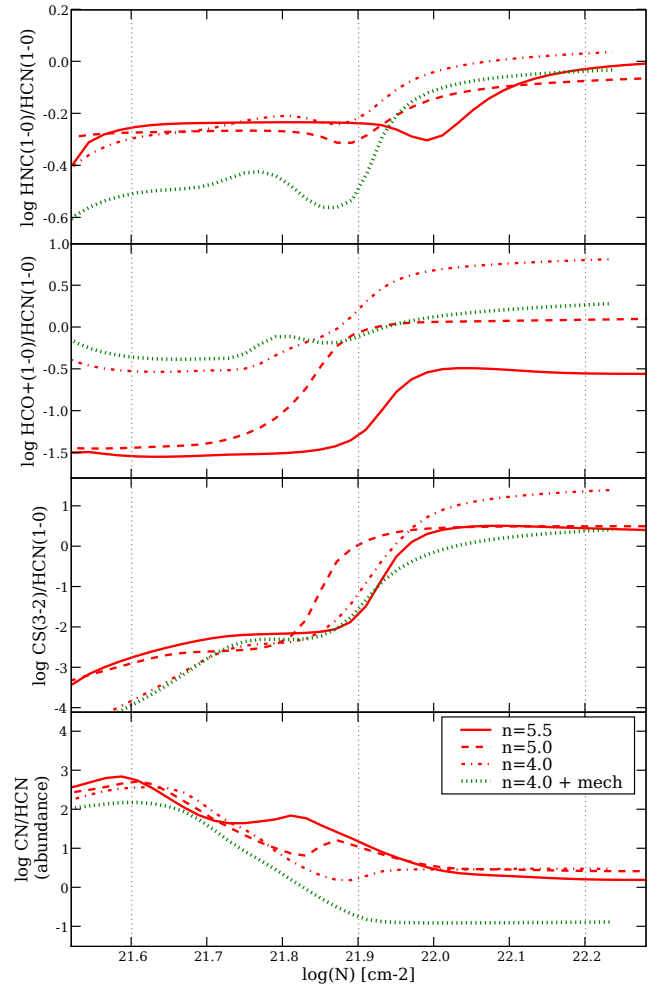}
\caption{Predicted line ratios for HNC\,(1$-$0)/HCN\,(1$-$0),
  \HCOP\,(1$-$0)/HCN\,(1$-$0), and CS\,(3$-$2)/HCN\,(1$-$0) and the
  abundance ratio of CN/HCN for varying environments. The four assumed
  conditions represent an average density ($n$ = 10$^{n}$ cm$^{-3}$)
  for the medium (red lines) and the addition of mechanical heating at
  the lowest density (green lines). For the CS diagram, a Sulfur
  depletion factor of 400 has been assumed. The vertical dotted lines
  designate the column density regions to be used in the
  discussion of the evolutionary diagrams in Section
  \ref{sec:sign-evol-molec}.  }\label{theoratios}
\end{figure}

The high-density component remaining after subsequent generations of
stars in the evolving starburst will decrease in mass and luminosity
and will be characterized by a systematically decreasing average
density and increasing temperature, although regions of relatively
higher density will remain. These changing conditions, as well as
variation of the cosmic ray flux density and the radiation fields,
will strongly influence the molecular chemistry.

Simulations using extensive physical and chemical modeling show that
the observed intensity ratios of high-density tracers HCN, HNC, and
\HCOP\, in (most) extra-galactic sources are well represented by an
environment that is dominated by Photon Dominated Regions (PDR) with
densities ranging from $\numd$ = 10$^{5.5}$ to 10$^{4}$ cm$^{-3}$
\citep{LoenenSBM2008}. Few sources in current sample show
characteristics suggesting X-ray Dominated Regions (XDR) (see section
\ref{hnc}). This density range confirms the densities anticipated for
star-formation regions but also suggests global evolutionary changes
in the nuclear environment during the course of an outburst.  Because
of different critical densities for the molecules, particularly the
intermediate density tracer lines \HCOP\,(1$-$0) and CN\,(1$-$0) (see
Table~\ref{crit}) may become thermalized during early stages of
evolution.

The observed intensity ratios for extra-galactic environments may
serve to diagnose any global change in the density and other
properties of the nuclear ISM. The expected variation of the HNC,
\HCOP, and CS line ratios with HCN have been calculated using
radiative transfer modeling within a physical/chemical network
\citep[see][]{LoenenSBM2008,LoenenBS2010,MeijerinkS2005}, while for
CN the abundance ratio is presented.  Fig.~\ref{theoratios} presents
the predicted line ratios for PDR environments with three
representative densities ($\numd$ = 10$^{5.5}$, 10$^{5.0}$, and
10$^{4.0}$ cm$^{-3}$). Variation of the UV-radiation fields only
leads to modest changes in the line ratios, because the UV flux is
largely attenuated at the high column densities where the molecules
are abundant.  In analogy with earlier simulations, the UV flux has
been taken to be 160 ergs s$^{-1}$ cm$^{-1}$ = 10$^{5}$ Habing units
\citep{LoenenSBM2008}, which is appropriate for starburst
environments.  An additional simulation curve considers the
effect of additional mechanical heating resulting from feedback at a
low rate of 5 x 10$^{-20}$ ergs s$^{-1}$ cm$^{-3}$ in the lowest
density environment ($\numd$ = 10$^{4}$) based on a SFR =  
2.5 \solmass yr$^{-1}$ \citep{LoenenSBM2008}. This heats the gas to a
temperature of approximately 200 K.

The predicted ratios in Fig.~\ref{theoratios} display significant
variation for a varying global density and for different column
densities representing the surface and interior regions of clouds from
which the observable emission emerges. For the ratios HNC/HCN,
\HCOP/HCN, and CS/HCN the cloud interiors dominate the observed
line ratios, while for CN the cloud surface would dominate the
observed line ratio.  The predicted values cover the range for the
observed ratios.

\begin{figure}[t!]
\includegraphics[width=9.0cm,clip]{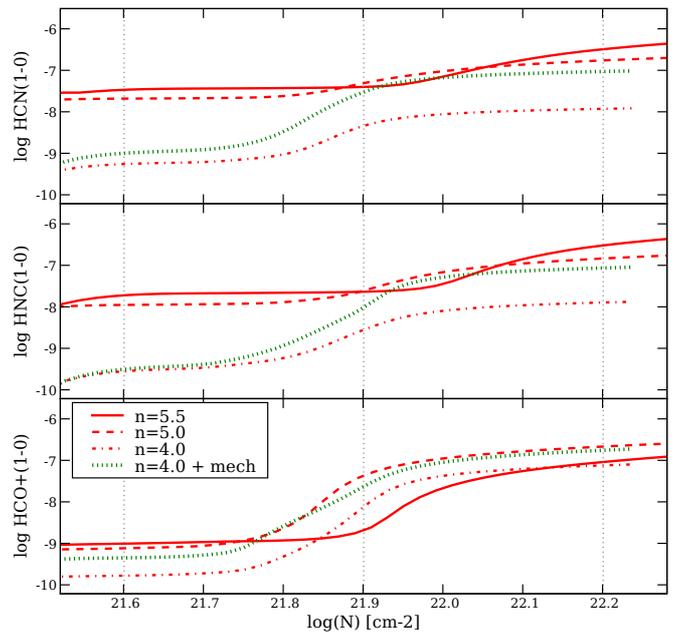}
\caption{Predicted line strength for HCN\,(1$-$0), HNC\,(1$-$0), and
\HCOP\,(1$-$0) for varying environments. The four assumed conditions
represent an average density ($n$ = 10$^{n}$ cm$^{-3}$) for the medium
(red lines) and the addition of mechanical heating at the lowest
density (green dotted lines). The vertical dotted lines designate the
column density regions that will be considered in the further discussion.}
\label{theorlines}
\end{figure}

\subsection{HCN as a molecular indicator}
\label{sec:indicators}

The HCN molecule has been identified as the best high-density
(HD) tracer to estimate the amount of high-density material and to
describe the state of the starburst \citep[see][]{GaoS2004a,
  BaanEA2008}.  Extensive modeling shows that of the prominent
HD-tracers, HCN is least sensitive to chemical and physical changes
and may best serve as an indicator of the high-density molecular
component \citep[see][]{Papadop2007, LoenenBS2007,
  LoenenSBM2008}.  The predicted line strength of emission lines of
HCN\,(1$-$0), HNC\,(1$-$0), and HCO$^{+}$ are presented in
Fig.~\ref{theorlines} for three densities in the range of 10$^{5.5}$
- 10$^{4}$ cm$^{-3}$ and for the effect of low-density mechanical
heating. The physical parameters of UV-flux and mechanical heating
used for the simulations are identical to those used for
Fig.~\ref{theoratios} (see Sect. \ref{sec:chem-evol-starb}).

The emission line strength in Fig.~\ref{theorlines} varies between the
cloud interior and the cloud surface and at higher densities. Furthermore, 
the changing ambient density varies from above to below the critical 
density, which results in different excitation states for all molecules.  
HCN(1$-$0) appears less sensitive to these changes,
because of its higher critical density (see Table~\ref{table1}), but is
also quite insensitive to additional physical effects.  Even though, the
destruction of HNC at higher temperatures would increase the HCN abundance,
this can at most only double the abundance.  Mechanical heating boosts
the line strength of all three molecules, but at lower densities (n(H)
= 10$^{4}$ cm$^{-3}$) this merely compensates for the effect of a
decrease in density. \HCOP might seem more stable in
Fig.~\ref{theorlines}, but this is simply because the ambient density is above 
n$_{crit}$ most of the time. More over, \HCOP is
inherently more sensitive to radiative/chemical effects, because it is
an ion. Additional ionization can enhance it and higher electron
abundance can destroy it.

The physical and chemical modeling does not incorporate any clumpiness
of the medium, which could affect the (integrated) strength of the emission 
lines. However, early PDR models for a clumpy medium \citep{Spaans1996, SpaansDishoeck1997} 
indicate that opaque clumps (A$_{V}$ $>$ 4 mag) only 
modestly affect the HCN/CO ratio, which also appears applicable for a large-scale 
nuclear ISM. In addition, the elemental abundance ratios, particularly C/N and O/N, 
among extragalactic sources would strongly affect the chemical 
balance. This would affect the strength of molecular emission lines 
from regions with lower column densities but not with high column densities.

Although all HD tracers will vary due to changing ISM conditions, it
appears that HCN is least perturbed by other physical and chemical
processes of the three molecular species considered here and may well
serve as a reliable indicator of the HD gas component in the
source. The HCN\,(1$-$0) (and higher) line strength does not necessarily
need to vary linearly with time or with FIR luminosity; it only needs
to be a monotone (smooth) function of the changing environment.

The validity of HCN as a representative HD tracer has been questioned
\citep{GraciaCarpioGPC2006, GraciaCarpioEA2008} particularly because
the HCN abundance can be enhanced in the highest luminosity objects
due to IR pumping \citep{AaltoSWH2007} and XDR chemical enhancement
\citep{LeppD1996}.  This XDR chemical enhancement is not depth
dependent and boosts HCN over a narrow range of ionization rates and
only for specific n$_H$ and F$_X$ \citep[see][]{MeijerinkS2005}. 
Also, the XDR component will typically have a much smaller annular scale
than the PDR, and will therefore suffer more from beam dilution \citep{SchleicherSK2010}.
Therefore, even a single AGN embedded in a starburst environment would
not likely cause a measurable boost of the HCN line strength.
Additionally, HCN enhancement at the highest FIR luminosities due to
IR pumping would still not preclude a smooth variation of the HCN
strength during a starburst.

\begin{figure}[t!]
\begin{center}
\includegraphics[width=8cm,clip]{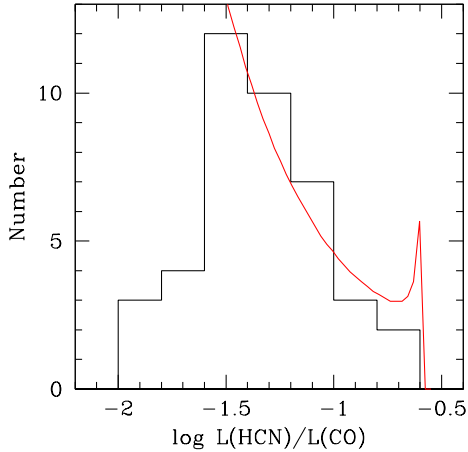}
\caption{A histogram of observed L(HCN)/L(CO) values versus the
  predicted values from simulations. The curve corresponds to the
  (red) prediction curve in Fig.~\ref{simul}c. } \label{histo}
\end{center}
\end{figure}

In the following discussion the ratio of the high-density and
low-density components of the ISM will be used to indicate the
sequence of evolutionary time based on the simulation of Section
\ref{sec:evolution}. The emission of the low-density component is well
represented by CO\,(1$-$0), which is less affected by a nuclear
starburst because it originates in a region much larger than the
starburst region.  Consequently, the HCN\,(1$-$0)/CO\,(1$-$0) ratio may
serve as a indicator of the relative HD depletion over time.

A simple test of the viability of HCN\,(1$-$0)/CO\,(1$-$0) as an indicator
of the (progressing) evolutionary time may be the comparison in
Fig.~\ref{histo} of the distribution of the ratios of the sources used
in the discussions below (no upper limits; see Table 2) with the shape
of the predicted curves in Fig.~\ref{simul}d (see
Sect. \ref{sec:evolution}). While the sample is incomplete at lower
$L_{HCN}$/$L_{CO}$ values, this simple model agrees quite well with
observations although the vertical scaling is still unknown.

\begin{figure}[t!]
\begin{center}
\includegraphics[width=8cm,clip]{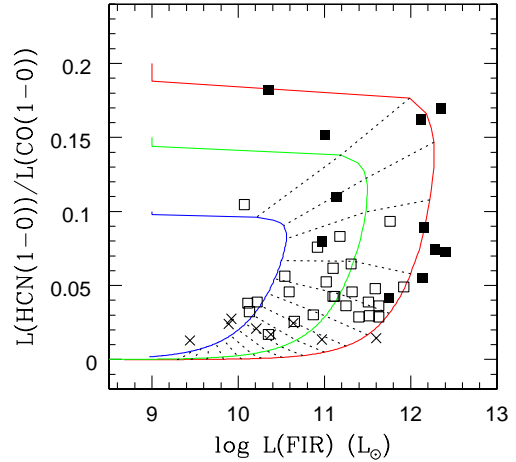}
\caption{Variation of the observed L(HCN)/L(CO) ratio with the
  FIR-luminosity predicted from simulations. The predicted starburst
  evolution curves for the HD/LD ratio are taken from
  Fig.~\ref{simul}c. The location of the data points may be used to
  estimate the evolutionary age of the starburst. Evolutionary ages
  are indicated with dotted lines starting from the top at timesteps
  2, 5,10, 20, 30, .. 140.  Each timesteps is of the order of 5 x 10$^{5}$
  yr. The filled sources symbols are for ULIRGs and MM sources,  the 
  open squares are LIRGs/starbursts and the crosses  represent evolved 
  starburst as discussed below. The crosses correspond to the grey 
  squares in Figs 6-9. } \label{HCN_CO-LFIR}
\end{center}
\end{figure}

A second test of the use of HCN\,(1$-$0)/CO\,(1$-$0) as evolutionary time
indicator is a comparison in Fig.~\ref{HCN_CO-LFIR} of the ratios
versus the FIR-luminosity with the predictions made in
Fig.~\ref{simul}c (see Sect. \ref{sec:evolution}).  The locations of
the data points are consistent with them following the predicted
evolutionary curves.  Assuming that the preliminary model in
Sect. \ref{sec:evolution} depicts the first-order evolutionary
behavior of a starburst, the location of data points relative to the
evolutionary curves would be an indication of the evolutionary age of
the starburst. Evolutionary ages are indicated by dotted lines covering 
a period of 10$^{8}$ yr. More detailed models need to be constructed to
accurately apply this concept to the evolution of a starburst.

\section{Signatures of Evolution of the Molecular ISM}
\label{sec:sign-evol-molec}

\begin{figure}[t!]
\includegraphics[width=7.5cm,clip]{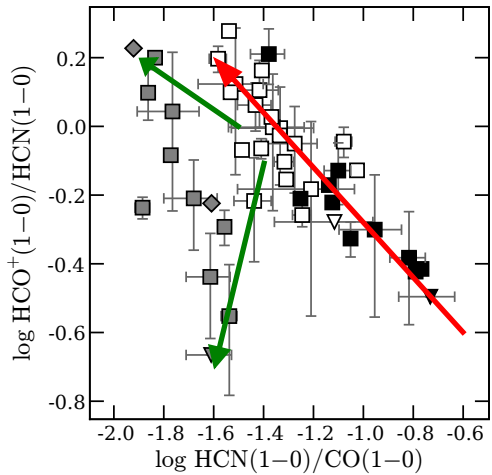}
\caption{The ratio of \HCOP\,(1$-$0)/HCN\,(1$-$0) versus the
  HCN\,(1$-$0)/CO\,(1$-$0).  The variation of \HCOP\, compared with
  the variation of the ratio of another high-density and a low-density
  tracer. The data points are squares for reliable values or triangles
  for upper and lower limits. Filled symbols indicate ULIRGs and OH MM 
  sources in the sample. The open symbols represent LIRGs/starbursts. 
  The sources with grey symbols are at later stages of starburst evolution 
  (corresponding to crosses in Fig.5 ). The red arrow designates the predicted effect
  of density variation and the green arrows designate the predicted
  effect of feedback through mechanical heating as simulated in
  Fig.~\ref{diagn1}. }
\label{diagn1}
\end{figure}

A molecular line ratio indicates changes in the chemical and physical
environment that may affect both molecules in a different manner.  The
line ratios for a number of characteristic molecular transitions have
been presented as a function of the HCN\,(1$-$0)/CO\,(1$-$0) line
ratio in Figs.~\ref{diagn1}$-$\ref{diagn4}. The line ratios for
high-density tracers in Table 2 are from \cite{BaanEA2008} with some
corrections of the calibration for some literature data. The
high-luminosity ULIRGs and OH Megamasers (OH MM), which are at early
stages of evolution, are located on the right-hand side of the
diagrams (filled symbols).  Evolved sources at lower $L_{FIR}$ (grey
symbols) are located on the left of the diagrams and may have been
affected by feedback during these late stages of evolution (see
section \ref{sec:HCOP} below).

The diagrams display a systematic move towards lower
$L_{HCN}$/$L_{CO}$ values in the diagram as a result of the evolution
of the starburst. Any spread of the observed line ratios (along
y-axis) would result from variation of environmental effects and
excitation conditions.
 
\subsection{The HCO$^+$/HCN - HCN/CO  relation}
\label{sec:HCOP}

The diagram of the \HCOP\,(1$-$0)/HCN\,(1$-$0) data
(Fig.~\ref{diagn1}) displays two well-defined distributions of data
points which are both decreasing along the HCN\,(1$-$0)/CO\,(1$-$0)
axis \citep[see][for a first version]{BaanEA2008}. The main
distribution contains ULIRGs and luminous starburst galaxies while the
second group consists of low FIR luminosity sources that have been
affected by feedback (see below). Both distributions suggests a steady
increase of the \HCOP\,(1$-$0)/HCN\,(1$-$0) ratio, while the strength
of both transitions decreases during the evolution of the outburst.

The \HCOP\,(1$-$0)/HCN\,(1$-$0) ratio in PDR-dominated environments is
predominantly sensitive to the mean density of the medium.  At higher
densities it decreases because of the increased dissociative
recombination rate of \HCOP\, with free electrons. At lower molecular
densities and at later stages of starburst evolution, the enhanced
ionization balance ($F_{UV}/n$) would enhance \HCOP, while at
$n$$\leq$10$^{4}$ cm$^{-3}$ the increased cosmic ray flux resulting from 
the SN of massive stars again destroys \HCOP.

Modeling of the predicted effect of changing the average density of
the ISM show a significant increase of the \HCOP\,(1$-$0)/HCN\,(1$-$0)
ratio for a change in density from 10$^{5.5}$ to 10$^{4}$ cm$^{-3}$
(Fig.~\ref{theoratios}).  This trend agrees with the general trend of
the data points as shown by the red arrow in Fig.~\ref{diagn1}.

Feedback by mechanical heating at a lower density of 10$^{4}$
cm$^{-3}$ lowers the \HCOP\,(1$-$0)/HCN\,(1$-$0) ratio at the cloud
centers (by a factor of three), but it is complicated by the mixed
effects of an increasing temperature and a decreasing density.
Feedback modeling results in Fig.~\ref{theoratios} suggest a
displacement of the \HCOP\,(1$-$0)/HCN\,(1$-$0) data points towards
lower values that depends on feedback intensity (in between two green
arrows ). This would move sources from the main distribution towards
the distribution of evolved sources (filled grey symbols at the left
side of Fig.~\ref{diagn1}).

Modeling also shows that higher \HCOP/HCN values result from the
presence of an AGN and the creation of dominant XDRs
\citep{LoenenSBM2008}. However, this change would be accompanied by an
increase in the HNC/HCN ratios, which is not observed in the data.  It
should be noted that the separation of the evolved sources from the
main body would indicates that the nuclear feedback conditions vary on
very short timescales.

The main distribution of data points in the \HCOP/HCN diagram is
consistent with the predictions for a systematic change of the line
ratio towards lower HCN/CO values resulting from environmental changes
during the evolution of a starburst.  A second group of data points
for evolved sources, which display the effects of feedback, displays a
different dependence for the \HCOP/HCN ratio on evolutionary time.

 \subsection{The HNC/HCN - HCN/CO  relation}\label{hnc}

\begin{figure}[t!]
\includegraphics[width=7.5cm,clip]{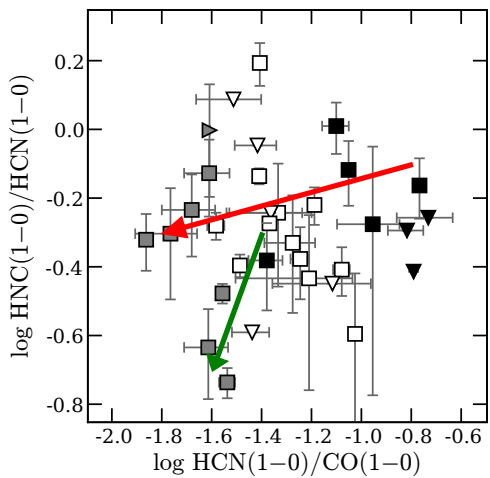}
\caption{The ratio of HNC\,(1$-$0)/HCN\,(1$-$0) versus the
HCN\,(1$-$0)/CO\,(1$-$0) ratio.  The ratio of HCN/CO represents the
progression of time for the nuclear activity running from right to
left.  The symbols and colors are as in Fig.~\ref{diagn1}.
} \label{diagn2}
\end{figure}

The HNC\,(1$-$0)/HCN\,(1$-$0) ratio displays data points covering a
large range of values (factor of 8; Fig.~\ref{diagn2}). However, it
should be noticed that data points for ULIRGs and also for evolved
galaxies are separated from the central distribution (forming three
bands of points). The distribution of data points displays a weak
overall decrease with decreasing HCN/CO by a factor of 1.6.

The HNC\,(1$-$0)/HCN\,(1$-$0) ratio is sensitive to the heating of the
environment \citep{LoenenSBM2008}. For values larger than unity, the
dominant heating source would be X-rays in XDRs.  NGC 7469, with the
largest ratio in our sample, is known to have a circum-nuclear
starburst and a black hole X-ray source.  Alternatively to X-ray
heating, the enhancement in these sources may be caused by pumping
with a very warm FIR radiation field that dominates over the
collisional processes for densities up to 10$^{6}$ cm$^{-3}$
\citep{AaltoSWH2007}. This process would operate at early stages of
the evolution with high densities and intense IR radiation fields,
which could explain the apparent HNC/HCN over-luminosity of the
distinct group of OH MM/ULIRG sources in Fig.~\ref{diagn2} (l.t.r.:
NGC 4418, Arp 220, IRAS15107+0724, and Mrk 231).

For HNC\,(1$-$0)/HCN\,(1$-$0) ratios in the range of 0.5 to unity, the
dominant heating source is the UV radiation in young PDRs. Ratios
smaller than 0.5 can not be explained by the steady state models of
PDRs (and XDRs). In a study of additional heating sources,
\cite{MeijerinkSI2006} found that cosmic ray heating is not sufficient
to influence the HCN and HNC abundances. On the other hand, an
increase in the temperature by means of SNe and SNR shocks will
decrease the HNC/HCN ratio, by opening an additional chemical
conversion path from HNC to HCN \citep{SchilkeEA1992}.  This
conversion of HNC at later stages of evolution of the starburst is
facilitated by turbulent heating of gas in the central and densest
regions of the ISM and is a temperature (T $\ge$ 100K) dependent
chemical process \citep{LoenenSBM2008}.

The simulations from Sect.~\ref{sec:chem-evol-starb} show that the
HNC/HCN ratio would systematically decrease with a factor 1.6 when the
average density drops from 10$^{5.5}$ to 10$^{4}$ cm$^{-3}$ as depicted 
with the red arrow. This arrow is representative of the average response of 
the emission regions with the highest column densities and agrees with the
observed average value for the ratio of around 0.6. With mechanical heating,
the HNC/HCN ratio may be reduced by as much as 2.5 at smaller HCN/CO
values as depicted with the green arrow. The onset of the effects of
mechanical feedback would vary with the type of galaxy and the
intensity of the starburst.  The general distribution of data points
is consistent with the changes expected for the line ratio as a result
of the lowering of density and the introduction of mechanical heating
from feedback. The reality of band structure in the data points
requires more data and further study.

The HNC/HCN diagram displays three apparent groups of data
points making a large (vertical) spread of the line ratio and possibly suggests 
enhanced ratios for the high-luminosity ULIRGs (see discussion above).  
The group of evolved sources at low HCN/CO forms a separate group 
of sources that shows little variation with time. 
While the red and green arrows depict the average time behavior of the 
distribution, the vertical spread suggests that additional physics affects the 
HNC/HCN line ratios. Nevertheless, the HNC/HCN
would also show a (weak) systematic variation with evolutionary time
until feedback changes the environment.

\subsection{The CN/HCN - HCN/CO  relation}
\label{sec:CN}

\begin{figure}[t!]
\vspace{-3mm}
\includegraphics[width=7.5cm,clip]{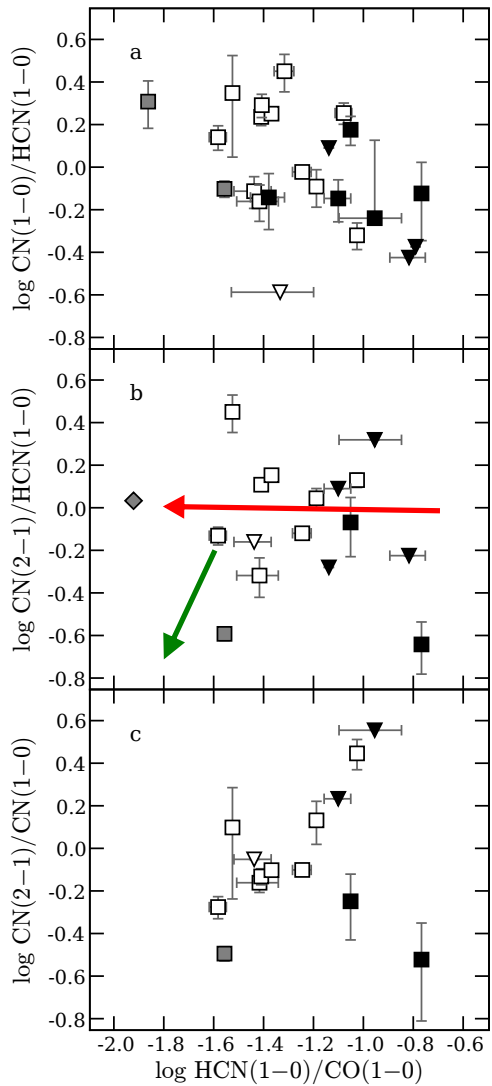}
\caption{The CN\,(1$-$0)/HCN\,(1$-$0) and CN\,(2$-$1)/HCN\,(1$-$0)
ratios versus the HCN\,(1$-$0)/CO\,(1$-$0) ratio. Frame c) presents
the CN\,(2$-$1)/CN\,(1$-$0) line ratio.  The symbols and colors are as in
Fig.~\ref{diagn1}.} \label{diagn3}
\end{figure}

The variation of the two CN/HCN ratios versus the HCN/CO ratio are
presented in Fig.~\ref{diagn3}.  The CN\,(1$-$0)/HCN\,(1$-$0) ratio
(Fig.~\ref{diagn3}a) displays a distribution that increases with
decreasing HCN/CO ratios.  On the other hand, the
CN\,(2$-$1)/HCN\,(1$-$0) distribution (frame b) shows twice the spread
of the CN\,(1$-$0)/HCN\,(1$-$0) ratio with a consistently zero slope.

The CN emission is generally enhanced in PDR-dominated Galactic
\citep{GreavesC1996, RodriguezEA1998} and extra-galactic
environments \citep{FuenteEA2005}. To a large extent, the CN is a
photo-dissociation product of HCN and HNC in the irradiated outer
layers of molecular clouds and serves as a diagnostic of the FUV and
cosmic-ray driven gas-phase chemistry
\citep[see][]{RodriguezEA1998,BogerS2005}.

The gradual increase in the CN\,(1$-$0)/HCN\,(1$-$0) ratio results
from the relatively low critical density of the CN\,(1$-$0) transition
(similar to \HCOP\,(1$-$0)). As the average density drops, the
CN\,(1$-$0) line strength at lower column densities will become
enhanced and less thermalized.   In addition, the CN\,(2$-$1) emitting
volume (towards cloud centers) reduces relative to the CN\,(1$-$0) 
emitting surface regions (low-column density) during the outburst.

The abundance simulations of PDR-dominated emission regions show that
the CN/HCN abundance ratio in regions with high column densities
remains close to unity when lowering the density from 10$^{5.5}$ to
10$^{4}$ (Fig.~\ref{theoratios}). Radiation transfer calculations have
not yet been done for CN.  This absence of variation of the CN
abundance is consistent with the CN\,(2$-$1)/HCN\,(1$-$0) data points in
Fig.~\ref{diagn3}b (red arrow).  The introduction of mechanical
heating due to feedback at later evolutionary times (green arrow))
would lower the relative abundance by as much as a factor of 20, which
is not yet seen in the data.

Inspection of the data points in Figs.~\ref{diagn3}ab shows that two 
luminous ULIRG sources, Arp 220 and Mrk 231, have an enhanced 
CN\,(1$-$0) transition and a reduced CN\,(2$-$1) transition. This also 
shows in the line ratios in Fig.~\ref{diagn3}c. This apparent lowering 
of the T$_{ex}$ may be result from peculiar pumping conditions in 
very luminous ULIRGs, possibly related to the anomalous
FIR pumping of HNC in these same sources \citep{AaltoSWH2007}. 

Assuming that the two ULIRG sources are indeed subject to anomalous 
conditions, the CN\,(1$-$0)/HCN trend would be more upwards and the 
CN\,(2$-$1)/HCN trend would be slightly downward with decreasing 
HCN/CO. In addition, the observed CN\,(2$-$1)/CN\,(1$-$0) ratio 
(Fig.~\ref{diagn3}c) displays a (nearly) linear dependence for the regular 
starbursts and LIRGs with HCN/CO and evolutionary age. Since the
abundance of CN remains relatively constant, this factor 10 variation
in the line ratio could be attributed to excitation effects or variation of the 
density.  Attributing this change to (excitation) temperature only suggests a 
change in effective T$_{ex}$ from 20 to 9 K, which is rather implausible.
Although this relation is still tentative because of the small number of
data points, this line ratio could be an indicator of the changing 
density and the difference in critical densities of the two lines. 

The CN/HCN diagrams suggest a systematic increase of the
CN\,(1$-$0)/HCN\,(1$-$0) ratio with evolutionary time accompanied by
no-change for the CN\,(2$-$1)/HCN\,(1$-$0) ratio.  This would result in a 
(density-related) systematic lowering of the CN\,(2$-$1)/CN\,(1$-$0) ratio 
with evolutionary time.  Detailed modeling of CN line ratios (in progress)
is required to explain the environmental influence on the ratios in
ULIRGs and the effect of feedback in evolved starbursts.

\begin{figure}[t!]
\includegraphics[width=7.5cm,clip]{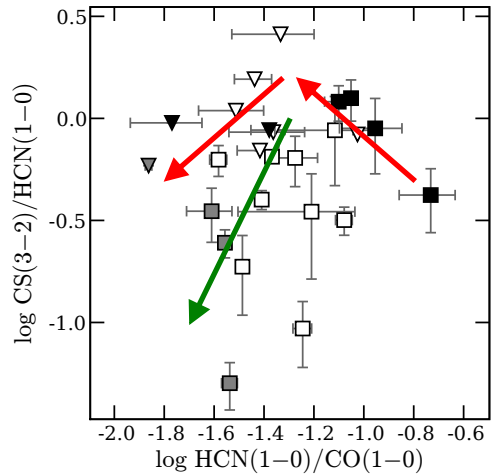}
\caption{The ratios of CS\,(3$-$2)/HCN\,(1$-$0) versus the
  HCN\,(1$-$0)/CO\,(1$-$0) ratio.  The symbols and colors are as in
  Fig.~\ref{diagn1}.} \label{diagn4}
\end{figure}

\subsection{The CS/HCN - HCN/CO  relation}

The CS\,(3$-$2)/HCN\,(1$-$0) diagram (Fig.~\ref{diagn4}) also displays
a large spread in data points (factor 30) and shows band-like
sub-structure similar to the HNC\,(1$-$0)/HCN\,(1$-$0) diagram of
Fig.~\ref{diagn2}. Again the OH MM/ULIRG sources form the upper
envelope with enhanced CS\,(3$-$2) emission.  The diagnostic diagram
of the CS\,(3$-$2)/HCN\,(1$-$0) ratio displays a downward trend going
to lower HCN/CO values during the course of an outburst.

The observed CS\,(3$-$2)/HCN\,(1$-$0) ratios mostly lie below unity,
which is slightly lower than predicted from modeling
(Fig.~\ref{theoratios}). These lower ratios may indicate that the
average Sulfur-depletion in the sources is somewhat higher than the
value of 400 used for Fig.~\ref{theoratios}c
\citep[see][]{LoenenBS2010}.

As a density tracer to first order with a high critical density, the
CS emissions depend strongly on the clumped high-density regions with
the highest column density in the ISM. The lower density (outer)
regions of molecular structures contribute little to the overall
emission (see also Fig.~\ref{theoratios}c).  CS may be enhanced in
photon-dominated regions due to reactions involving S$^{+}$
\citep{SternbergD1995}, while an enhanced cosmic ray flux would
deplete its abundance.  The grouping in the diagram may be caused by
environmental changes related to feedback and a changing PDR
population.

The modeling results show that the {\it average
CS\,(3$-$2)/HCN\,(1$-$0) line ratio at high column density} would
first increase when the average density drops to 10$^{5}$ cm$^{-3}$
and then decrease again (or flatten) when it reaches 10$^{4}$
cm$^{-3}$ (Fig.~\ref{theoratios}).  The introduction of mechanical
heating and feedback at 10$^{4}$ cm$^{-3}$ would lower the ratio by a
factor up to 10.  These theoretical tendencies have been indicated in
Fig.~\ref{diagn4} with two red arrows (for density) and a green arrow
(for feedback).

The CS/HCN diagram also displays three apparent groups of data points making 
a large (vertical) spread of the line ratio, similar to the HNC/HCN diagram. 
Again there is the suggestion of enhanced ratios for the high-luminosity ULIRGs.  
The group of evolved sources at low HCN/CO also shows little variation with 
time.  While the red and green arrows depict the average time behavior of the 
distribution, the vertical spread suggests that additional physics affects the 
CS/HCN line ratios and results in a (weak) systematic variation with
evolutionary time until feedback changes the environment.

\section{Summary}
\label{sec:phenomenon}

The evolutionary stages of a nuclear starburst represent a
well-determined sequence of events that would affect the ISM in a
complex but predictable manner.  As a result, the characteristics of
the ISM and the subsequent star-formation process will change
systematically during the course of the outburst and ultimately leads
to the termination of the process. The dominant effects describing the
nuclear ISM are the steady depletion of the high-density molecular
component, the decreasing average density of the medium, the
increasing temperature resulting from the heating due to feedback.

A  simple  model has  been  discussed  that  describes the  decreasing
high-density component of the ISM during an outburst. It has also been
argued that HCN is the molecular species that is least affected by the
chemical  and  physical changes  in  the ISM  and  best  serves as  an
indicator of the  high-density component in the ISM.  For this reason,
all molecular  variations in this  study are measured relative  to the
strength  of   the  HCN\,(1$-$0).  Molecular species will react 
differently to  the changing environment of the nuclear  ISM and these 
molecules will  reveal different aspects of the changing physical and 
chemical environment. In this example,  the  decreasing
HCN\,(1$-$0)/CO\,(1$-$0)   ratio   serves  as   an   initial  measure   of
evolutionary  time for the  starburst. Systematic changes may be seen 
for all characteristic line ratios with
evolutionary time. The observed ratios with HCN\,(1$-$0) for the two
transitions with lower critical densities, CN\,(1$-$0) and
\HCOP\,(1$-$0), display a systematic increase with evolutionary time
of the starburst. Furthermore, the CN\,(2$-$1)/CN\,(1$-$0) ratio for starbursts and LIRGs 
(tentatively) decreases
systematically with evolutionary time. All these ratios may be used as
indicators of evolutionary time of the starburst.

The two other transitions with higher critical densities, HNC\,(1$-$0)
and CS\,(3$-$2) display a more complicated picture because it appears
that additional physical processes affect the ratios, particularly for 
the highest-luminosity ULRGs. Both transition display weak evolutionary 
variation with evolutionary age. While FIR pumping may
account for the anomalous HNC/HCN ratio in these ULIRGs, a pumping
mechanism for CS has yet to be identified. Besides HNC and CS, CN 
also displays anomalous excitation during the
early ULIRG stage. Here the CN\,(1$-$0) line is enhanced and the
CN\,(2$-$1) line is reduced relative to HCN\,(1$-$0), which may reduce the
CN\,(2$-$1)/CN\,(1$-$0) ratio during early stages of evolution.

In addition to the main group of sources that display a systematic
increase in the observed ratios with a lowering of the HCN/CO ratio, a
group of evolved sources has been identified that would display a
different dependence on evolutionary time. Besides the systematic 
lowering of density in the PDR-dominated environment (from 
10$^{5.5}$ to 10$^{4}$ cm$^{-3}$), the feedback from SNe and SNRs 
during later stages of evolution provides mechanical heating leads 
to lower ratios for all line ratio with respect to HCN\,(1$-$0) 
\citep{LoenenSBM2008}.

The observed variation of characteristic line ratios may be used to
diagnose the evolutionary state of the nuclear activity in
galaxies. This would allow to establish an evolutionary sequence for
ULIRGs and (future) SMGs and a way to understand and diagnose their 
physical processes.  The systematic changes of the characteristic
molecular line ratios already provide a first view of evolutionary
sequence for the nuclear activity. While the relation between the
HCN\,(1$-$0)/CO\,(1$-$0) line ratio and evolutionary time only
provides relative time during the evolution of the starburst, further
modeling of the observed changes would provide a more accurate
translation to evolutionary time. In addition, sequential modeling to
reconstruct the changing nuclear environment will provide a better
understanding and identification of the different physical and
chemical processes that determine the evolution of the molecular
environment during a nuclear outburst. Additional observational data
of multiple molecular transitions will further improve the
understanding of the complex myriad of competing processes.

%

\bibliographystyle{aa}
\bibliography{LIBMolecV4}
%
%
\onecolumn
\small

\begin{longtable}{@{\extracolsep{-3mm}}lllllllll@{\extracolsep{-3mm}}}
\caption{Transitional line ratios for high-density tracer molecules}\\
\hline 
\hline
Source &  & $\frac{{\rm HCN\,(1-0)}}{{\rm CO\,(1-0)}}$ & $\frac{{\rm HNC\,(1-0)}}{{\rm HCN\,(1-0)}}$ & $\frac{{\rm HCO^+\,(1-0)}}{{\rm HCN\,(1-0)}}$ & $\frac{{\rm CN\,(1-0)}}{{\rm HCN\,(1-0)}}$ & $\frac{{\rm CN\,(2-1)}}{{\rm HCN\,(1-0)}}$ & $\frac{{\rm CN\,(2-1)}}{{\rm CN\,(1-0)}}$ & $\frac{{\rm CS\,(3-2)}}{{\rm HCN\,(1-0)}}$ \\
IRAS & name &  &  &  &  &  &  &  \\
\hline
00450$-$2533	& NGC\,253	& -1.41 (-2.768)	& -0.13 (-1.406)	& -0.06 (-1.234)	& 0.24 (-0.805)		& 0.11 (-1.129)		& -0.13 (-1.190)	& -0.39 (-1.366) \\
01053$-$1746	& IC\,1623	& -1.44 (-2.249)	& \ldots		& 0.06 (-0.731)		& \ldots		& \ldots		& \ldots	        & \ldots         \\
01403$+$1323	& NGC\,660	& -1.34 (-1.778)	& -0.24 (-0.649)	& -0.00 (-0.502)	& $<$-0.58		& \ldots		& \ldots	        & $<$0.42        \\
01484$+$2220	& NGC\,695	& -1.84	                & \ldots		& 0.20			& \ldots		& \ldots		& \ldots	        & \ldots         \\
02193$+$4207	& NGC\,891	& -1.77 (-2.325)	& -0.30 (-0.749)	& 0.05 (-0.267)		& \ldots		& \ldots		& \ldots	        & \ldots         \\
02401$-$0013	& NGC\,1068	& -1.08 (-2.196)	& -0.41 (-1.197)	& -0.04 (-1.038)	& 0.26 (-0.683)		& \ldots		& \ldots	        & -0.49 (-1.297) \\
03317$-$3618	& NGC\,1365	& -1.21 (-1.521)	& -0.43 (-0.709)	& -0.18 (-0.422)	& \ldots		& \ldots		& \ldots	        & -0.45 (-0.727) \\
03419$+$6756	& IC\,342       & -1.56 (-2.800)	& -0.47 (-1.658)	& -0.29 (-1.221)	& -0.10 (-1.154)	& -0.59 (-1.747)	& -0.49 (-1.606)	& -0.61 (-1.409) \\
04315$-$0840	& NGC\,1614	& -1.53 (-2.784)	& \ldots		& \ldots		& 0.35 (0.051)		& 0.45 (-0.245)		& 0.10 (-0.167)	        & \ldots         \\
05059$-$3734	& NGC\,1808	& -1.25 (-2.315)	& -0.37 (-1.000)	& -0.26 (-1.376)	& -0.02 (-1.292)	& -0.12 (-1.282)	& -0.10 (-1.345)	& -1.03 (-1.474) \\
05083$+$7936	& VII\,Zw\,31	& -1.31	                & \ldots		& -0.15	                & \ldots	        & \ldots		& \ldots		& \ldots         \\
05414$+$5840	& \ldots	& -1.44 (-2.211)	& $<$-0.59		& -0.21 (-0.690)	& -0.11 (-0.878)	& $<$-0.16		& $<$-0.05	        & $<$0.20        \\
06106$+$7822	& NGC\,2146	& -1.87 (-3.381)	& -0.32 (-1.045)	& 0.10 (-0.674)		& 0.31 (-0.289)		& \ldots		& \ldots	        & $<$-0.23       \\
07160$-$6215	& NGC\,2369	& -1.37 (-1.841)	& $<$-0.24		& \ldots		& \ldots		& \ldots		& \ldots	        & $<$-0.06       \\
09293$+$2143	& NGC\,2903	& -1.62 (-2.316)	& -0.63 (-1.166)	& -0.44 (-0.906)	& \ldots		& \ldots		& \ldots	        & \ldots         \\
09517$+$6954	& M\,82	        & -1.59 (-2.690)	& -0.28 (-1.321)	& 0.20 (-0.860)		& 0.14 (-0.736)		& -0.13 (-1.144)	& -0.27 (-1.196)	& -0.20 (-0.961) \\
09585$+$5555	& NGC\,3079	& -1.61 (-2.299)	& -0.13 (-0.723)	& $<$-0.66		& \ldots		& \ldots		& \ldots		& -0.45 (-0.979) \\
10257$-$4338	& NGC\,3256	& -1.42 (-2.142)	& $<$-0.04		& 0.11 (-0.549)		& -0.16 (-0.869)	& -0.31 (-0.992)	& -0.16 (-1.157)	& $<$-0.15       \\
11143$-$7556	& NGC\,3620	& -1.12 (-1.490)	& $<$-0.45		& $<$-0.28		& \ldots		& \ldots		& \ldots	        & -0.05 (-0.387) \\
11176$+$1351	& NGC\,3628	& -1.68 (-2.301)	& -0.23 (-0.803)	& -0.21 (-0.740)	& \ldots		& \ldots		& \ldots	        & \ldots         \\
11257$+$5850a	& Arp\,299a	& -1.38 (-2.190)	& -0.38 (-0.925)	& 0.21 (-0.523)		& -0.14 (-0.670)	& \ldots		& \ldots	        & $<$-0.05       \\
11257$+$5850bc	& Arp\,299bc	& -1.32 (-2.361)	& \ldots		& -0.10			& 0.45 (-0.244)		& \ldots		& \ldots	        & \ldots         \\
11506$-$3851	& \ldots	& -0.74 (-1.333)	& $<$-0.25		& $<$-0.49		& \ldots		& \ldots		& \ldots	        & -0.37 (-0.833) \\
12112$+$0305	& \ldots	& -1.13	                & \ldots		& -0.22			& \ldots		& \ldots		& \ldots	        & \ldots         \\
12243$-$0036	& \ldots	& -1.10 (-2.017)	& 0.01 (-0.756)		& -0.13 (-0.816)	& -0.14 (-0.794)	& $<$0.09		& $<$0.24	        & 0.09 (-0.620)  \\
12540$+$5708	& Mrk\,231	& -0.77	                & -0.16 (-0.860)	& -0.41			& -0.12 (-0.518)	& -0.64 (-1.199)	& -0.52 (-0.833)	& \ldots         \\
12542$+$2157	& NGC\,4826	& -1.89 (-3.065)	& \ldots		& -0.23 (-1.374)	& \ldots		& \ldots		& \ldots	        & \ldots         \\
13025$-$4911	& NGC\,4945	& -1.37 (-3.114)	& -0.27 (-1.863)	& 0.03 (-1.675)		& 0.26 (-1.257)		& 0.16 (-1.496)		& -0.10 (-1.647)	& -0.18 (-1.823) \\
13126$+$2452	& \ldots	& $<$-1.92	        & \ldots		& $>$0.23		& \ldots		& $>$0.04		& \ldots	        & \ldots         \\
13183$+$3423	& Arp\,193	& -1.54	                & \ldots		& 0.28			& \ldots		& \ldots		& \ldots	        & \ldots         \\
13341$-$2936	& M\,83	        & -0.98 (-1.483)	& -0.53 (-1.030)	& -0.35 (-0.640)	& \ldots		& \ldots		& \ldots	        & -1.09 (-1.491) \\
13428$+$5608	& Mrk\,273	& -0.79	                & $<$-0.41		& -0.42			& $<$-0.37		& \ldots		& \ldots	        & \ldots         \\
15065$-$1107	& \ldots	& -1.77 (-2.269)	& \ldots		& \ldots		& \ldots		& \ldots		& \ldots	        & $<$-0.02       \\
15107$+$0724	& \ldots	& -0.96 (-1.510)	& -0.27 (-0.440)	& -0.30 (-0.651)	& -0.24 (-0.114)	& $<$0.32		& $<$0.56	        & -0.04 (-0.441) \\
15327$+$2340	& Arp\,220	& -1.05 (-2.440)	& -0.12 (-0.786)	& -0.32 (-1.259)	& 0.18 (-0.628)		& -0.06 (-0.574)	& -0.24 (-0.711)	& 0.10 (-0.538)  \\
16504$+$0228	& NGC\,6240	& -1.03	                & -0.59 (-0.894)	& -0.13			& -0.32 (-1.161)	& 0.13 (-0.981)		& 0.45 (-0.340)	        & $<$-0.07       \\
17208$-$0014	& \ldots	& -1.14	                & \ldots		& -0.17			& $<$0.09		& $<$-0.28	        & \ldots	        & \ldots         \\
20338$+$5958	& NGC\,6946	& -1.49 (-3.484)	& -0.39 (-1.516)	& -0.07 (-1.874)	& \ldots		& \ldots	        & \ldots	        & -0.72 (-1.098) \\
21453$-$3511	& NGC\,7130	& -1.19 (-2.470)	& -0.22 (-1.121)	& \ldots		& -0.09 (-0.786)	& 0.05 (-0.906)	        & 0.14 (-0.505)	        & \ldots         \\
22025$+$4205	& \ldots	& -0.82 (-1.611)	& $<$-0.29		& -0.38 (-0.821)	& $<$-0.42		& $<$-0.22	        & \ldots	        & \ldots         \\
22347$+$3409	& NGC\,7331	& $>$-1.61           	& \ldots		& $<$-0.22		& \ldots		& \ldots	        & \ldots	        & \ldots         \\
23007$+$0836	& NGC\,7469	& -1.41	                & 0.20 (-0.650)	        & 0.17			& 0.30 (-0.607)		& \ldots		& \ldots	        & \ldots         \\
23134$-$4251	& NGC\,7552	& -1.28 (-1.917)	& -0.33 (-0.754)	& -0.05 (-0.597)	& \ldots		& \ldots	        & \ldots	        & -0.19 (-0.747) \\
23156$-$4238	& NGC\,7582	& -1.52 (-2.051)	& $<$0.09		& 0.13 (-0.216)		& \ldots		& \ldots	        & \ldots	        & $<$0.04        \\
23365$+$3604	& \ldots	& -1.26	                & \ldots		& -0.21			& \ldots		& \ldots	        & \ldots	        & \ldots         \\
23488$+$1949	& NGC\,7771	& -1.34	                & \ldots		& -0.02			& \ldots		& \ldots	        & \ldots	        & \ldots         \\
23488$+$2018	& Mrk\,331	& -1.54	                & \ldots		& 0.10			& \ldots		& \ldots	        & \ldots	        & \ldots         \\
\hline
\end{longtable}

\end{document}